\documentclass[prl,reprint,superscriptaddress,showpacs]{revtex4-1}
\usepackage{graphicx}
\usepackage{amsmath}
\usepackage{dcolumn}

\begin{document}

\title{Direct measurement of the electron energy relaxation dynamics in metallic wires}

\author{Edouard Pinsolle}
\author{Alexandre Rousseau}
\author{Christian Lupien}
\author{Bertrand Reulet}
\affiliation{D\'{e}partement de Physique, Universit\'{e} de Sherbrooke, Sherbrooke, Qu\'{e}bec J1K 2R1, Canada.}
\date{\today}

\begin{abstract}
We present measurements of the dynamical response of thermal noise to an ac excitation in conductors at low temperature.
From the frequency dependence of this response function - the (noise) thermal impedance - in the range $1$ kHz-$1$ GHz we obtain direct determinations of the inelastic relaxation times relevant in metallic wires at low temperature: the electron-phonon scattering time and the diffusion time of electrons along the wires. 
Combining these results with that of resistivity provides a measurement of heat capacity of samples made of thin film.	
The simplicity and reliability of this technique makes it very promising for future applications in other systems. 

\end{abstract}

\maketitle
\indent Energy relaxation of electrons in a conductor is a very important issue both on an applied and fundamental point of view. 
For example, the energy relaxation rate determines the bandwidth of hot electron bolometers used to detect electromagnetic radiation through heating of the electron gas \cite{burke_mixing_1999}.
Hence, we can distinguish between devices where relaxation processes are due to electron-phonon coupling (phonon cooled \cite{kawamura_low_1997}) and devices where the electron gas cools down by out diffusion of hot electrons in reservoirs (diffusion cooled \cite{skalare_large_1996}).
In the former devices, the relevant inelastic time is the electron-phonon scattering time $\tau_{e-ph}$ while in the latter it is the diffusion time $\tau_D$.\\
\indent On a fundamental point of view, inelastic times are key parameters for example for quantum correction to electron transport, electron localization at low temperature \cite{Li_two_2012} and non-equilibrium effects \cite{kawamura_low_1997}.
Usual transport measurements such as conductance versus temperature fail to give access to the energy relaxation dynamics. 
At high temperature the conductance of a metal is usually determined by electron-phonon interaction, but $\tau_{e-ph}$ is so short that it is not accessible through transport and one has to use ultrafast optical methods to measure it \cite{rogier_femtosecond_1995}. 
At low temperature in disordered conductors, the conductance is determined by the elastic mean free path, and independent of electron-phonon interactions. 
As a consequence, inelastic times are not determined directly. 
They are obtained as fitting parameters in measurements of weak localization corrections to conductance \cite{santhanam_inelastic_1984,pieper_measurement_1992} (in normal metals or superconductors above $T_c$) or in tunneling experiments which access the energy distribution function \cite{Energy_pothier_1997}.
A direct determination of inelastic times would provide a way to better test theories of quantum transport at low temperature.\\
Directly observing energy relaxation requires monitoring the energy or temperature of the electrons.
This is precisely what we have done by measuring the amplitude of the temperature oscillation of metallic wires heated by an ac Joule power. 
This is a measurement of the frequency dependent thermal impedance between the electron gas and the relevant thermal reservoirs, here phonons and electrical contacts. 
To measure the electron temperature on a short time scale we have used Johnson noise detected at high frequency, i.e. we have measured the noise thermal impedance introduced in \cite{reulet_noise_2005}. 
As we show below, from these measurements we deduce the temperature-dependent electron-phonon scattering time as well as the diffusion time in samples of different lengths and made of different materials. 
Combining these results with that of conductance, we deduce the heat capacity of the samples, which would be completely undetectable using conventional methods because of their small value. 
In the following we discuss the theoretical expectations for the frequency dependence of the thermal impedance in the limits of phonon cooling and diffusion cooling. 
Then we describe the experiment and discuss the results: thermal impedance, relaxation rate and heat capacity.\\
\indent We consider conducting wires heated by Joule power oscillating at frequency $f$ with an amplitude $\delta P_J(f)$, which induces a modulation of the electron temperature at the same frequency $\delta T_e(f)$. 
The thermal impedance $R(f)$ is defined by $\delta T_e(f)=R(f)\delta P_J(f)$. 
$R(f)$ is a complex quantity since at finite frequency there is a phase shift between the power and temperature oscillations.
At zero frequency, $R(f=0)$ is simply the inverse of the usual thermal conductance $G_{th}$.\\
\indent The frequency dependence of $R(f)$ has been calculated for a metallic wire in different regimes \cite{reulet_noise_2005}. For long enough samples the energy relaxation of the electron gas is dominated by electron-phonon interactions. This occurs when $L\gg L_{e-ph}$ where $L_{e-ph}$ is the electron-phonon scattering length given by $L_{e-ph}^2=D\tau_{e-ph}$ with $D$ the diffusion coefficient. The electron temperature $T_e$ is uniform along the sample and obeys:
\begin{equation}
	C_e \frac{\partial T_e}{\partial t}=P_J(t)-P_{e-ph}
	\label{heat1}
\end{equation}
where $C_e$ is the heat capacity of the electron gas, and $P_{e-ph}$ is the electron-phonon cooling power. The thermal impedance $R(f)$ is given by:
\begin{equation}
	R(f)=\frac{G_{e-ph}^{-1}}{1+2i\pi f \tau_{e-ph}}
\label{NTI_eph}
\end{equation}
where $G_{e-ph}=\frac{\text{d}P_{e-ph}}{\text{d}T_e}$ is the thermal conductance between electrons and phonons.\\
\indent For shorter samples $L\ll L_{e-ph}$ electron-phonon processes are inefficient and the energy relaxation is dominated by diffusion of hot electrons into the contacts. 
This cooling mechanism is more important close to the contacts so the temperature $T_e(x)$ is not uniform along the sample  even though the ac heating is. The local electron temperature $T_e(x)$ \footnote{Here we suppose the electron electron interactions to be strong enough that a local temperature can be defined} obeys the heat diffusion equation:
\begin{equation}
	C_e \frac{\partial T_e(x,t)}{\partial t}=P_J(t)+\frac{\partial}{\partial x}(G_{WF}(x,t)\frac{\partial T_e(x,t)}{\partial x})
	\label{heat2}
\end{equation}
where $G_{WF}$ is the heat conductance related to the electrical conductance through the Wiedmann-Franz law. 
This equation has been solved in \cite{reulet_noise_2005} to give the full frequency dependence of $R(f)$. Here $R$ measures the response of the average temperature of electrons along the wire to the ac heating. We have checked that the frequency dependence of $|R(f)|^2$ is extremely well approximated by a Lorentzian decay with a characteristic frequency $\simeq 10.01/\tau_D$. As a result, both for $L\gg L_{e-ph}$ and $L\ll L_{e-ph}$, we expect a response function of the form:
\begin{figure}
\center
	\includegraphics[width=0.9\columnwidth]{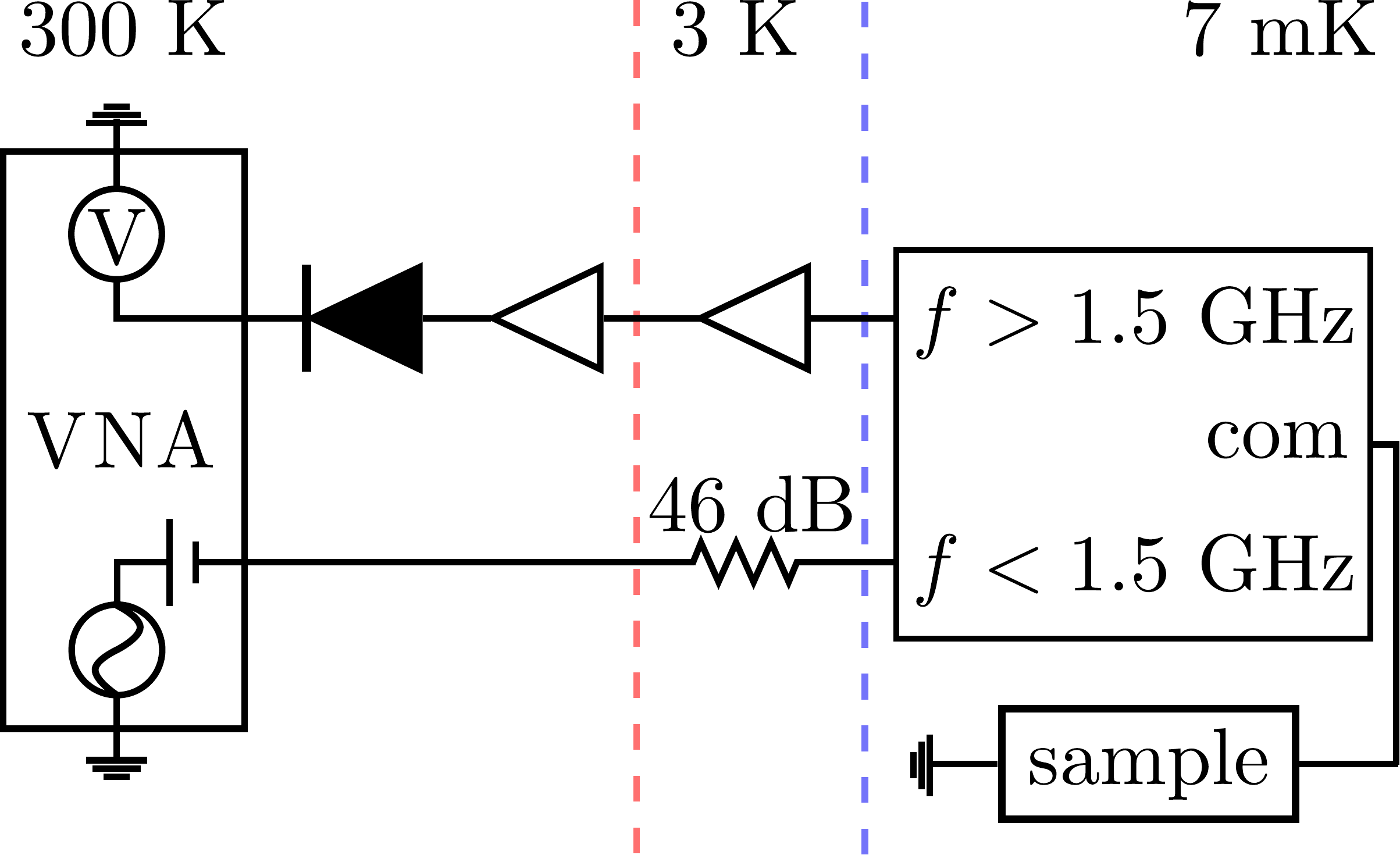}
	\caption{Experimental setup. Diode symbol represents a power detector. VNA=Vector Network Analyser}
	\label{Fig.1}
\end{figure}
\begin{equation}
	|R(f)|^2=\frac{R(0)^2}{1+(2\pi f/\Gamma(T_e))^2}
\label{NTI_bis}
\end{equation}
where $\Gamma$ the energy relaxation rate depends on the relaxation process.
The frequency dependence of $R(f)$ is a direct probe of $\Gamma$ without any assumption about $C_e$ as in previous work \cite{gershenson_millisecond_2001}. 
On the contrary, $C_e$ can be determined by combining measurements of $R(f)$ at low and high frequency.\\
\indent In the presence of several relaxation processes, the fastest relaxation usually dominates.
Since $\tau_{e-ph}$ is strongly temperature dependent and diverges at low temperature whereas $\tau_D$ is temperature independent, the energy relaxation is dominated by electron-phonon coupling at high temperature ($\tau_{e-ph}\ll\tau_D$) and diffusion at low temperature ($\tau_{e-ph}\gg\tau_D$).
Our measurement allows  continuous monitoring of $\Gamma$ as a function of temperature.\\
\indent We have measured $\Gamma(T_e)$ for six samples made of different metals (Al, Ag) and different geometries (the Al has been kept in its normal state with the help of a strong Neodymium permanent magnet).
The wires have length $L$ ranging from 5 $\mu \text{m}$ to 300 $\mu \text{m}$ and thickness $d$ of 10 nm for the shortest and 20 nm for the others, see Table \ref{table1}. 
The width has been adjusted to obtain a resistance of the order of 50 $\Omega$ for impedance matching purpose. 
\begin{table*}
	\begin{tabular}{c c c c c c c c c}
  \hline
  \hline
  Sample & $L$  & $w$ & $d$  &$R$& $A$ & $n$ & $D$ & $\tau_{D}$\\
	 & ($\mu$m) & ($\mu$m) &(nm)& ($\Omega$)& ($10^8 \text{s}^{-1}\text{K}^{-n}$) & &($10^{-3}$  m$^2$/s) &($\mu$s) \\
  \hline
  1(Al) & 310  & 20.7 & 20 & 41.0  & $1.8  \pm 0.1$   &~2.60 $\pm$ 0.05 &~5.2   $\pm$ 0.3   & 66  $\pm$ 6\\
  2(Al) & 53.4 & 5.6  & 20 & 134   & $1.72 \pm 0.03 $ &~2.89 $\pm$ 0.04 &~1.01  $\pm$ 0.07  & 2.85  $\pm$ 0.08\\
  3(Al) & 25.6 & 2.63 & 10 & 104   & $3.65 \pm 0.04 $ &~2.88 $\pm$ 0.03 &~2.8   $\pm$ 0.3   & 0.233  $\pm$ 0.003\\
  4(Al) & 13.4 & 2.9  & 10 & 57.3  & $6.04 \pm 0.09 $ &~2.76 $\pm$ 0.05 &~2.4   $\pm$ 0.4   & 0.0562   $\pm$ 0.0006\\
  5(Al) & 5.36 & 0.35 & 10 & 80.0  & $5.70 \pm 0.37 $ &~2.87 $\pm$ 0.09 &~5.7   $\pm$ 0.7   & 0.0069  $\pm$ 0.0001\\
  6(Ag) & 50   & 1.5  & 15 & 28.61 & $1.50 \pm 0.06 $ &~3.40 $\pm$ 0.10 &~29.0  $\pm$ 6.0   & 0.134  $\pm$ 0.002\\
  \hline
  \hline
\end{tabular}
\caption{Sample parameters. $L$ is the wire length, $d$ is the thickness, $w$ is the width, $R$ the resistance, $A$ and $n$ are the parameters of the Eq. (\ref{equg}). $D$ is the diffusion coefficient calculated with $\sigma=n(E_F)e^2D$ where $\sigma$ is the conductivity, $e$ the electron charge  and $n(E_F)$ the density of state at Fermi level. $\tau_{D}$ is the diffusion time extracted from $R(f)$ measurements.}
\label{table1}
\end{table*}
The contacts, made of the same metal as the wires, are much larger($\text{400}\ \mu \text{m} \times \text{400}\ \mu \text{m}$) and thicker (200 nm) to make sure they behave as electron reservoirs (for discussion on non perfect reservoirs see \cite{henny_1/3_1999}).
Samples from 5 $\mu$m to 25 $\mu$m have been made by e-beam lithography and the metal has been deposited by double angle evaporation \cite{reese_niobium_2007,dolan_offset_1977}. 
In this process we first evaporate the wire followed by the contacts without breaking the vacuum, thus preventing the growth of an oxide at their interface. 
Samples 1 and 2 have been made in two photo-lithography steps.
A first one to make the wire and a second one for the contacts. 
The native oxide that develops on the wire between the two processes has been removed by ion milling before evaporating the contacts.\\
The experimental setup is presented in Fig. \ref{Fig.1}.
The sample, placed at the 10 mK stage of a dilution refrigerator, is dc and ac biased through the low frequency port of a diplexer by a time dependent voltage $V=V_0+\delta V \cos(2\pi f t)$ with $\delta V<V_0$. 
The dc part $V_0$ is used to control the sample mean electron temperature through a constant Joule heating $P_J=GV_0^2$ and allowed us to work between $\sim 50$ mK and $\sim 2$ K.
The superimposed ac power at frequency $f$, $\delta P_J(t)=2GV_0\delta V \cos(2\pi ft)$ modulates the electron temperature of the sample. 
To detect this temperature, we measure the rms amplitude of the voltage fluctuations (Johnson noise) generated by the sample. Indeed, the noise spectral density of voltage fluctuations $S_V$ is related to the electron temperature by $S_V=4k_BT_e/G$.
The voltage fluctuations are measured in the frequency band $\Delta F \simeq 1.5-5$ GHz (high frequency port of the diplexer) and amplified by a cryogenic amplifier placed at the 3 K stage of the dilution refrigerator. Their rms amplitude is detected by a power meter (diode symbol in Fig. \ref{Fig.1}) whose response time $\tau_{det}\sim1$ ns limits the maximum frequency at which the noise modulation can be detected, $f\lesssim1$ GHz. 
The detected power $P_{det}$ contains the noise generated by the sample and the amplifier.
Its oscillation at frequency $f$, $\delta P_{det}(f)$, detected with a vector network analyzer (VNA), is given by:
\begin{equation}
	\delta P_{det}(f)=\eta(f)R(f)\delta P_J(f) 
\end{equation}
where $\eta$ is the response function of the detection chain (which frequency dependence is dominated by the response time of the power detector) and $\delta P_J(f)$ the ac Joule power dissipated in the sample (note that due to imperfections and attenuation in the excitation line, the ac voltage across the sample is not known). 
Both need to be calibrated to extract $R(f)$.
In order to determine $\Gamma(T_e)$, it is enough to know the frequency dependence of $R(f)$, not its absolute value. Thus we consider the normalized thermal impedance:
\begin{equation}
	\frac{R(f)}{R(0)}=\frac{\eta (0)}{\eta (f)}\frac{\delta P_J(0)}{\delta P_J(f)}\frac{\delta P_{det}(f)}{\delta P_{det}(0)}=\Lambda(f) \delta P_{det}(f)
\end{equation}
At high temperature the frequency dependence of $R(f)$ is given by the electron-phonon scattering rate which increases with temperature ($\Gamma(T_e)=\tau^{-1}_{e-ph}=A\times T_e^n$ with $n\simeq3$) \cite{lawrence_calculation_1978}.
Thus for $T_e\gtrsim3$ K $R(f)$  is frequency independent below 1 GHz and the observed frequency dependence of $\delta P_{det}(f)$ reflects only that of the setup, which allows the determination of $\Lambda(f)$.\\ 
\indent Experiments have been performed at a phonon temperature of 10 mK. 
To make the link between the applied bias $V_0$ and the electron temperature we first measured the noise of the sample at low frequency ($1.5$ GHz) at equilibrium (no bias) as a function of the temperature $T_e$, as well as the noise at base temperature as a function of $V_0$.
From these two measurements we deduce the link between applied voltage and electron temperature, i.e. how much voltage $V_0$ is needed for the sample to generate as much noise as when it is at equilibrium at temperature $T_e$.
We verified that even on the shortest sample, heating by applying a dc voltage or by increasing the overall temperature of the dilution refrigerator leads to the same $\Gamma(T_e)$.\\
\begin{figure}
\center
	\includegraphics[width=1\columnwidth]{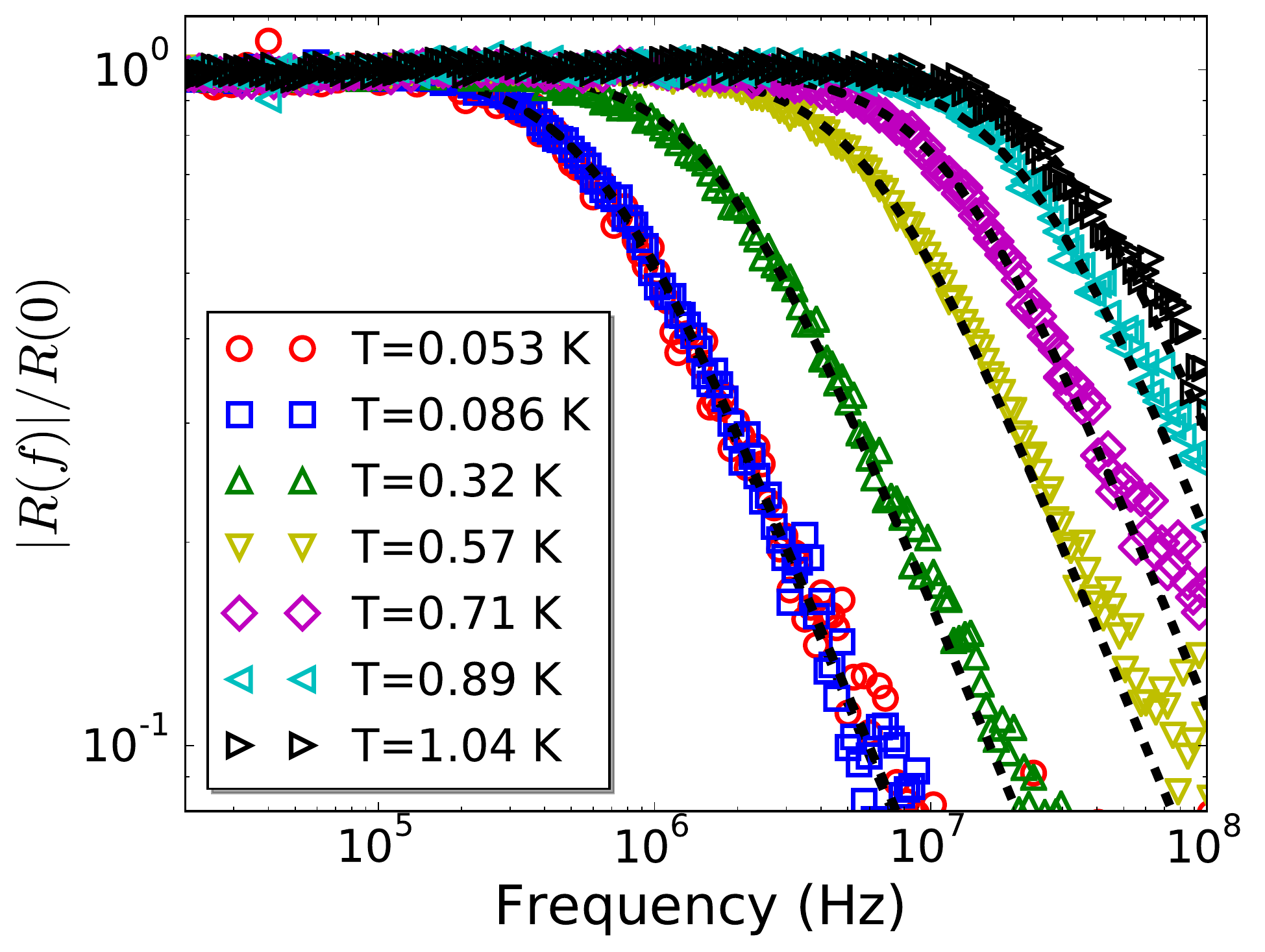}
	\caption{Amplitude of the normalized thermal impedance as a function of frequency for sample 2. The symbols are the experimental data and the dashed lines are fits according to Eq. (\ref{NTI_eph}). The different curves correspond to different electron temperatures from $\simeq$50 mK to $\simeq$1 K.}
	\label{Fig.2}
\end{figure}
\begin{figure}[t]
\begin{center}
	\includegraphics[width=1\columnwidth]{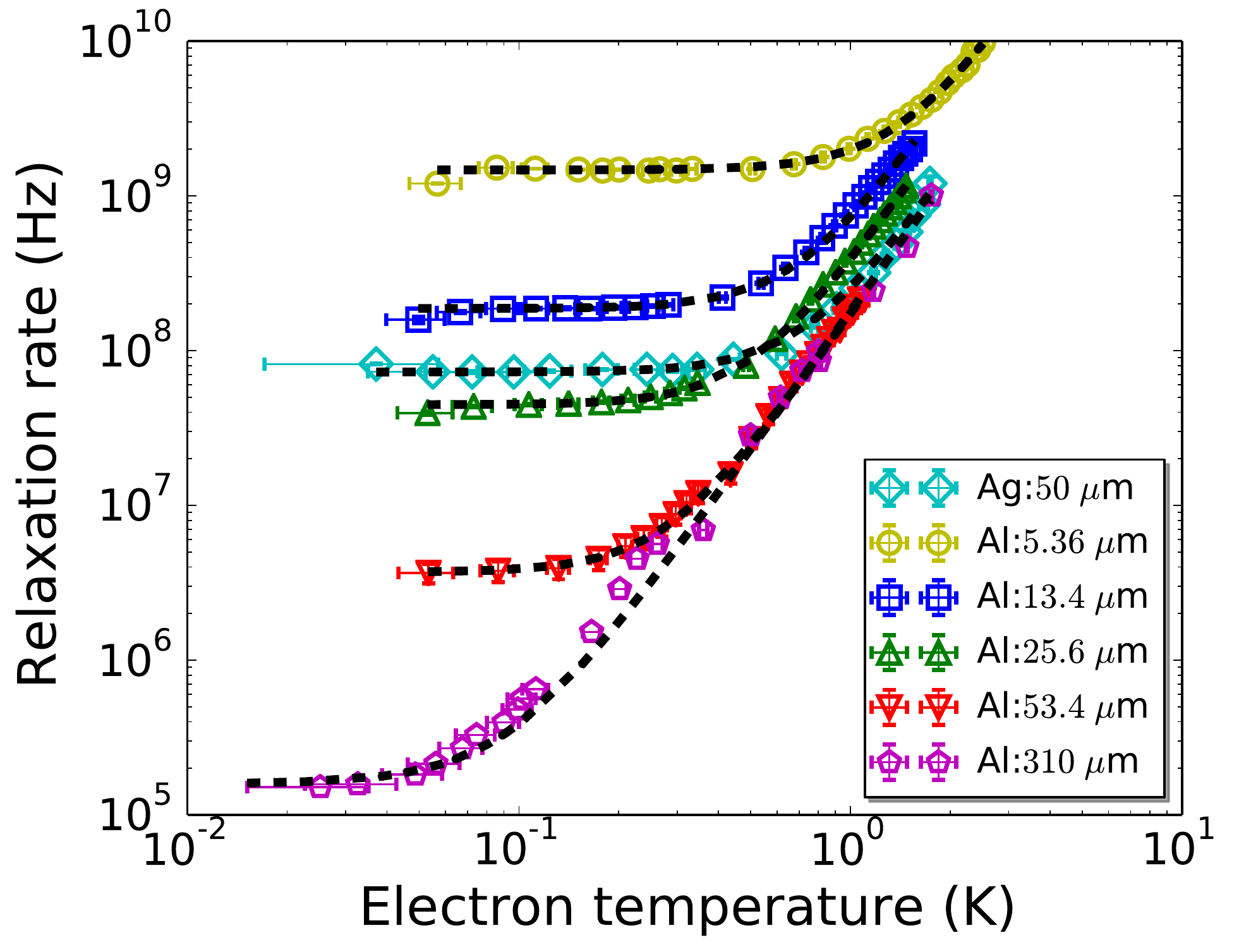}
	\caption{Energy relaxation rate as a function of electron temperature for all the samples. Dashed lines are fits according to Eq. (\ref{equg})}
	\label{Fig.3}
\end{center}
\end{figure}
In Fig. \ref{Fig.2}, we present the normalized thermal impedance versus frequency for sample 2 for electron temperatures between $53$\ mK and $1.04$\ K.
The symbols are the experimental data and the black dashed lines the fits according to Eq. (\ref{NTI_bis}).
The frequency dependence of $|R(f)|^2$ is very well fitted by a Lorentzian, $\Gamma(T_e)$ being the only fitting parameter.
We have performed this experiment for all the samples of Table 1 and extracted $\Gamma(T_e)$ on 5 orders of magnitude.\\
We present in Fig. \ref{Fig.3} the measured relaxation rates as a function of electron temperature for all the wires. 
At low temperature we observe a plateau, the relaxation rate does not depend on temperature. 
In this limit only diffusion cooling occurs, and $\Gamma(T_e)\simeq10.01/ \tau_D$.
At high temperature the observed $T_e^n$ dependency is characteristic of an electron-phonon cooling process \cite{lawrence_calculation_1978}.
In \cite{huard_electron_2007} the dynamic has only been calculated in the electron-phonon cooling or diffusion cooling regimes and not during the crossover.
We thus assume that the frequency dependence of $|R(f)|^2$ follows a Lorentzian even during the crossover between the two regimes with a relaxation rate given by the sum of the relaxation rates of the two processes:
\begin{equation}
	\Gamma(T_e)\simeq \frac{10.01}{\tau_D}+\text{A}T_e^n. 
	\label{equg}
\end{equation}
Dashed lines in Fig. \ref{Fig.3} are fits according to Eq. (\ref{equg}).
The parameters $A$, $n$ and $\tau_D$ extracted from the fits are summarized in table \ref{table1}.\\
\indent The plateau observed in $\Gamma(T_e)$ at low temperature, see Fig. \ref{Fig.3}, provides a direct determination of the diffusion time $\tau_D$ as a function of sample length.
We expect $\tau_D=L^2/D_{E}$ where $D_{E}$ is the energy diffusion coefficient,  since our experiment probes energy relaxation.
On the other hand the Einstein relation $\sigma=\kappa e^2D_{q}$ relates the conductivity $\sigma$ to the charge diffusion coefficient $D_{q}$, with  $\kappa$ the electronic compressibility, which reduces to the density of state at Fermi energy $n(E_F)$ for non interacting electrons.
At low temperature, since diffusion is the only way energy can be relaxed, one expects $D_q=D_E=D$ according to the Wiedemann-Franz law.
Thus, taking $n(E_F)$ in the free electron approximation, one deduces $D$ from the conductivity. Accordingly, we plot the product $D\tau_D$ vs. $L$ on the inset of Fig. \ref{Fig.4} for all the samples. One clearly observes $D\tau_D=L^2$ (solid line) except for the longest Al wire \footnote{Measurements on this sample required a high ac excitation to get measurable heating because of its large volume, which might cause the deviation we observe. 
This sample also shows an electron-phonon power law temperature dependence that significantly differs from all the other Al samples, see Table \ref{table1}.}.\\
\indent At high temperature the relaxation is dominated by electron-phonon interaction \footnote{In thin films below 1K, the Kapitza thermal resistance between the phonons of the wire and that of the substrate can be neglected \cite{wellstood_1994_hot}.} and $\Gamma(T)=AT_e^n$. We find $n\simeq 3$ (see Table \ref{table1}), the expected value for three dimensional phonon bath in the clean metallic limit \cite{lawrence_calculation_1978,santhanam_inelastic_1984}.
In previous experiments, $\tau_{e-ph}^{-1}$ has been reported to behave as $T_e^n$ with $n$ ranging from 2 to 4 depending on the nature of the disorder \cite{sergeev_electron_2000,bergmann_nonequilibrium_1990,bergmann_inelastic_1982,peters_1985,echternach_electron_1992,gershenson_millisecond_2001,roukes_hot_1985}. 
Disordered gold wires have been observed to behave as $T_e^{2.9}$ below 1K \cite{echternach_electron_1992}. As far as we know, electron-phonon relaxation rates of Al and Ag have not been measured below 1K, a temperature range hardly explored \cite{echternach_electron_1992,gershenson_millisecond_2001,roukes_hot_1985}.\\
\indent While the frequency dependence of the ratio $R(f)/R(0)$ provides a calibration-free method to determine the energy relaxation rate of the electrons, the measurement of $R(f)$ with absolute units contains more information. At low frequency, $R(0)= d T_e/d P_J$ is deduced from the voltage dependence of the temperature $T_e$.
In the phonon cooled regime $R(0)=G_{e-ph}^{-1}$. Combining this measurement with that of the scattering time $\tau_{e-ph}$ one can determine the electronic heat capacity $C_e=G_{e-ph}\tau_{e-ph}$.
We show on Fig. \ref{Fig.4} (red symbols) the heat capacity of sample $2$ in which electron-phonon dominates above 0.3 K, see Fig. \ref{Fig.3}. 
In the diffusion cooled regime the electronic heat capacity is determined by $C_e=G_{th}\tau_{D}$.
$G_{th}$ can be determined from $R(0)$ (i.e., calibrated dc noise measurement) or from the electrical conductance $G$ using the Wiedmann-Franz law.
Green symbols on Fig. \ref{Fig.4} show $C_e$ in the diffusion cooled regime obtained by combining thermal impedance measurements and conductance measurements.
We observe that these two limits coincide and are in good agreement with the free electron model (dashed line) \cite{Ashcroft}.
Note that for our shortest sample (sample 5) at 50 mK we have been able to detect an extremely small value of the heat capacity $\sim2.10^{-19}$ J.K$^{-1}$ which is orders of magnitude lower than what is done using usual thermodynamic techniques \cite{garden_thermo_2009}.
We have determined $C_e(T_e)$ in two limits. A theory for the full temperature dependence of $R(f)$ is required to extract $C_e(T_e)$ in the whole temperature range.\\
\begin{figure}
\center
	\includegraphics[width=1\columnwidth]{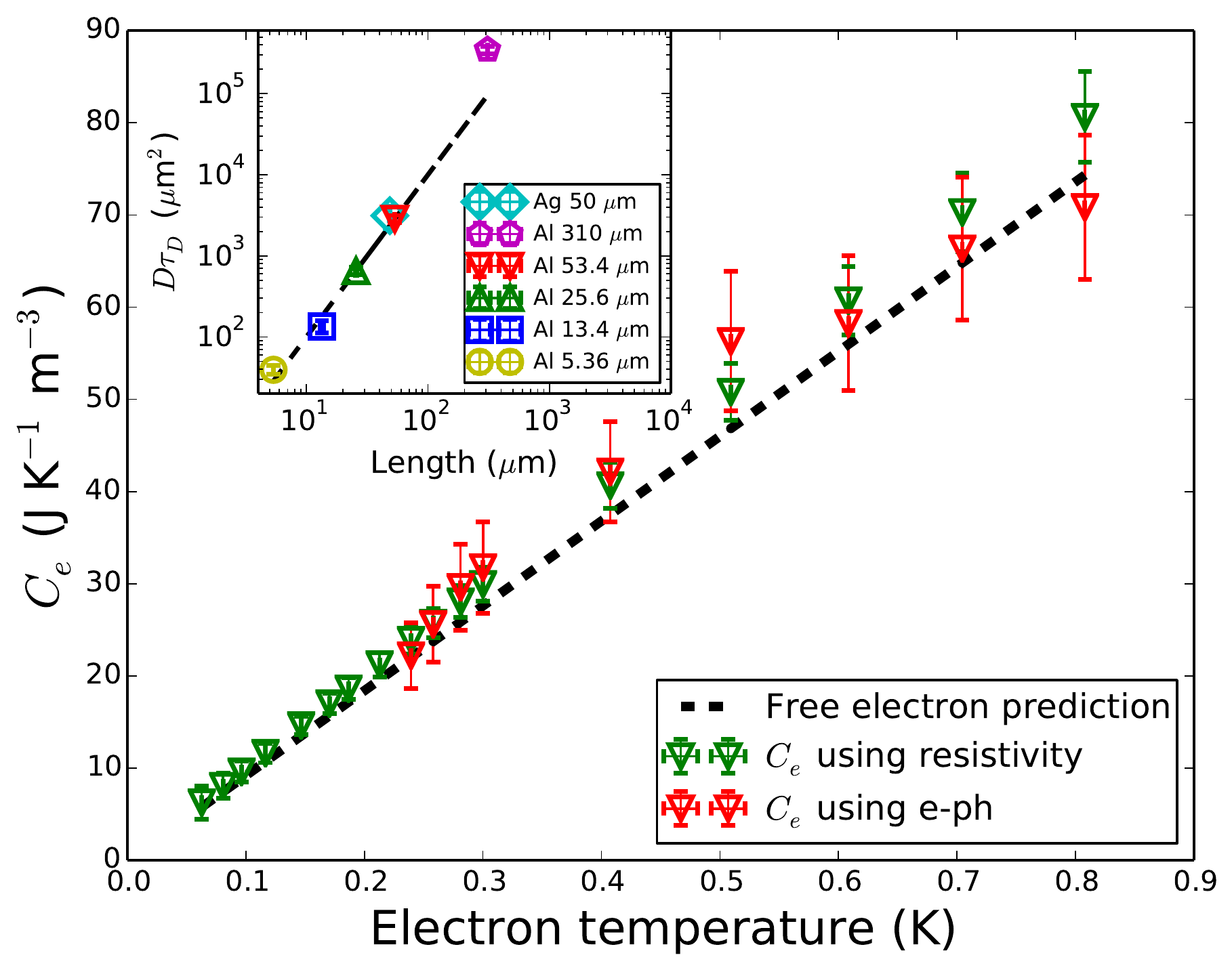}
	\caption{Electronic heat capacity of sample 2 as a function of temperature. Red triangles are obtain from $\Gamma(T_e)$ and $R(0)$ in the phonon cooled regime. Green triangles are obtain from $\tau_D$ and $G$ in the diffusion cooled regime.
Dashed line is the expected value for the free electron gas. Inset: product of the diffusion coefficient $D$ with the diffusion time $\tau_D$ as a function of wire length. The symbols are the experimental data, the dashed line is $L^2$.}
\label{Fig.4}
\end{figure}
\indent We have demonstrated a sub-kelvin direct measurement of inelastic times in wires made of simple metals, which provides the determination of the electron-phonon scattering time, the diffusion time and the electron heat capacity of the sample. Our approach is however extremely versatile, and of great interest to study interactions and electron diffusion in modern materials.
The measurement of the relaxation rate as a function of temperature or magnetic field could give a strong insight into the electron-phonon mechanism in graphene \cite{betz_supercollision_2012} and topological materials \cite{spivak_magneto_2016}.
The direct determination of the diffusion time would play a pivotal role in the study of transport in materials such as quasi-crystals or thin films with fractal geometry, for which a deviation from a quadratic scaling of $\tau_D$ with the length of the sample is expected \cite{roche_fermi_1998,sokolov_fractal_2011}.
Finally, our technique will allow the measurement of the electron heat capacity where conventional techniques simply do not work, in particular in samples which do not exist in bulk (nanowires, disordered thin films, 2D electron gas in heterostructures with strongly correlated materials).\\ 
\indent We are very grateful to M. Shen, D. Prober and R. Shoelkopf with whom preliminary experiments have been performed at Yale university \cite{these}. We also acknowledge fruitful discussion with D. Prober, M. Aprili, J. Gabelli, P. Fournier, R. Nourafkan, A-M Tremblay and technical help of G. Lalibert\'{e}. This work was supported by the Canada Excellence Research Chair program, the NSERC, the MDEIE, the FRQMT via the INTRIQ, the Universit\'{e} de Sherbrooke via the EPIQ, and the Canada Foundation for Innovation.


\bibliographystyle{apsrev4-1}
\bibliography{../biblio/All_library}

\end{document}